\documentclass[10pt,letterpaper]{article}
\usepackage[top=0.85in,left=1.6in,footskip=0.75in]{geometry}

\usepackage{amsmath,amssymb}

\usepackage{changepage}
\usepackage{subfig}
\usepackage[utf8x]{inputenc}

\usepackage{textcomp,marvosym}

\usepackage{cite}

\usepackage{nameref,hyperref}

\usepackage{microtype}
\DisableLigatures[f]{encoding = *, family = * }

\usepackage[table]{xcolor}

\usepackage{array}

\usepackage{caption,tabularx,booktabs}

\newcolumntype{+}{!{\vrule width 2pt}}

\newlength\savedwidth

\usepackage[aboveskip=1pt,labelfont=bf,font=small,labelsep=period,singlelinecheck=off]{caption}

\makeatletter
\renewcommand{\@biblabel}[1]{\quad#1.}
\makeatother

\usepackage{lastpage,fancyhdr,graphicx}
\usepackage{epstopdf,soul}
\begin{document}
\vspace*{0.2in}

\begin{flushleft}
{\Large
\textbf\newline{\textbf{Intentional binding enhances hybrid BCI control}}}
\newline
\\
Tristan Venot\textsuperscript{1},
Arthur Desbois\textsuperscript{1},
Marie Constance Corsi\textsuperscript{1},
Laurent Hugueville\textsuperscript{1},
Ludovic Saint-Bauzel\textsuperscript{2}
Fabrizio De Vico Fallani\textsuperscript{1*}
\\
\bigskip
\textbf{1} Sorbonne Université, Institut du Cerveau - Paris Brain Institute - ICM, CNRS, Inria, Inserm, AP-HP, Hôpital de la Pitié Salpêtrière, F-75013, Paris, France
\\

\textbf{2} Sorbonne Université, Institut des Systèmes intelligents et de robotiques ISIR, F-75005, Paris, France

\bigskip

* Corresponding author: fabrizio.de-vico-fallani@inria.fr

\end{flushleft}
\bigskip

\section*{Abstract}

\begin{small}
Mental imagery-based brain-computer interfaces (BCIs) allow to interact with the external environment by naturally bypassing the musculoskeletal system.

Making BCIs efficient and accurate is paramount to improve the reliability of real-life and clinical applications, from open-loop device control to closed-loop neurorehabilitation.
By promoting sense of agency and embodiment, realistic setups including multimodal channels of communication, such as eye-gaze, and robotic prostheses aim to improve BCI performance.
However, how the mental imagery command should be integrated in those hybrid systems so as to ensure the best interaction is still poorly understood.

To address this question, we performed a hybrid EEG-based BCI experiment involving healthy volunteers enrolled in a reach-and-grasp action operated by a robotic arm.
Main results showed that the hand grasping motor imagery timing significantly affects the BCI accuracy as well as the spatiotemporal brain dynamics.
Higher control accuracy was obtained when motor imagery is performed just after the robot reaching, as compared to before or during the movement. 
The proximity with the subsequent robot grasping favored intentional binding, led to stronger motor-related brain activity, and primed the ability of sensorimotor areas to integrate information from regions implicated in higher-order cognitive functions.

Taken together, these findings provided fresh evidence about the effects of intentional binding on human behavior and cortical network dynamics that can be exploited to design a new generation of efficient brain-machine interfaces.


\end{small}


\newpage

\section*{Introduction}

By establishing direct communication pathways between the brain and external devices, noninvasive brain-computer interfaces (BCIs) have the potential to become one of the most disruptive technology of the century.

BCIs interpret and translate neural signals, typically recorded with functional neuroimaging such as electroencephalography (EEG), into commands that can be understood by computers or devices. 
Humans communicate their intention by voluntarily modulating brain activity thanks to mental imagery tasks, such as hand grasping or mathematical operations \cite{wolpaw_brain-computer_2012}.
The potential impact of BCIs is huge, from understanding brain mechanisms underlying motor/cognitive control and neurofeedback, to the development of prototypes for substitution and rehabilitation of lost functions (e.g., stroke recovery)\cite{silvoni_brain-computer_2011,mane_bci_2020}.

A crucial requirement for BCIs is the usability in terms of performance, that is the ability to correctly detect and recognize the user's mental intent and ensure a reliable communication. 
On the computer side, the challenge is to correctly classify the user's intent. Basic research in electrophysiology as well as methodological advances in signal processing and machine learning have made considerable progresses \cite{lotte_review_2018}.
On the human side, the challenge is to learn to generate accurate brain signals so as to ease their detectability from the computer \cite{corsi_functional_2020,wander_distributed_2013}. Recent works on cognitive science demonstrated that the level of subjects' embodiment, i.e. the feeling of the environment as an extension of the own body, plays a crucial role in facilitating the BCI skill acquisition process \cite{serim_revisiting_2023}. To this end, the design of immersive setups including hybrid communication channels and bodily-like robotic effectors, can provide significant benefit in terms of BCI performance 
as compared to traditional display-based configurations \cite{cao_brain-actuated_2021,xu_continuous_2022}. 
For example, teleoperated androids and head-mounted displays led to easiness in motor imagery training and higher levels of embodiment resulting in an overall BCI performance improvement \cite{alimardani_brain-computer_2018,juliano_embodiment_2020,leeb_walking_2006,petit_integrated_2015,martens_towards_2012}. 
More in general,  embodiment could be even obtained through non-human looking robots \cite{aymerich-franch_non-human_2017,segil_measuring_2022}.  

These findings demonstrate the beneficial effect of considering multimodal setups in terms of classification accuracy but ignore how they eventually affect the brain dynamics. Yet, a better understanding of the brain organizational mechanisms associated with embodying BCIs would allow more biologically-grounded design optimization. 
A particularly delicate question is how the mental imagery process should be integrated in such sophisticated realistic setups in order to favor the feeling of the user to be in control, also called sense of agency \cite{haselager_did_2013,van_acken_tracking_2012,jeunet_chapter_2016,haggard_sense_2017}. 
Here, we hypothesized that the timing of the mental imagery process, i.e., the moment when subjects send the mental imagery command, plays a fundamental role in determining the overall BCI performance.  In particular, the delay between the mental intention and the related action might significantly affect the user's sense of agency,  a key prerequisite for embodiment.
To test this hypothesis, we realized a longitudinal randomized hybrid BCI experiment involving a group of healthy volunteers controlling a robotic arm to reach and grasp a target object. We tested three different mental imagery timing strategies: \textit{i)} before, \textit{ii)} during, and \textit{iii)} after the reaching phase.
We evaluated the BCI performance in terms of classification accuracy and sensitivity, and we investigated the brain dynamics in terms of motor-related activity and functional connectivity.
See \textbf{Methods} for details of the sample, experimental paradigm, and methods of analysis.

\section*{Results}

\subsection*{Behavioral performance}

Fifteen healthy right-handed subjects (8 females) participated in a randomized longitudinal EEG study consisting in controlling the reach-and-grasp action of a robotic arm via an hybrid-BCI (\textbf{Fig. 1a}). 
The goal was to use the eye-gaze to select a target object and grasp it by means of a right-hand motor imagery (MI). Across sessions, subjects were instructed to perform the MI task in different moments, i.e., \textit{S1)} before, \textit{S2)} during and \textit{S3)} after the reaching phase (\textbf{Fig. 1b}, \textbf{S1a}).
Each session started with a calibration, where subjects were instructed through a visual cue prompted on the table monitor to perform several trials of MI (grasp) and resting state (no-grasp) tasks. At this stage the robotic arm reached the target and the grasping depended on the given cue and not on the recorded brain activity (neurofeedback off).
After selecting the most relevant controlling EEG channels in terms of discriminant power spectra (\textbf{File S1}), subjects performed the same task but the robotic hand action was now controlled by the brain activity (neurofeedback on). Based on the input controlling features, a linear discriminant classifier determined the type of action, i.e. grasp/no grasp. Two consecutive control blocks were then realized to allow subjects to practice and to learn the task (\textbf{Fig. S1b}, \textbf{Methods}). 

To assess the role of the intrinsic subjects' motivation on their ability to control the BCI, we first measured their reward/effort ratio via an online questionnaire before the experiment (\textbf{Methods})\cite{pessiglione_why_2018}.
Results showed that the highest classification accuracies (correct/total trials) tended to be reached by the most motivated subjects (\textbf{Fig 2a}).
Then, we investigated how subjects became proficient and whether one timing strategy gave better performance.
In average subjects exhibited a significant learning effect across the control blocks regardless of the timing strategy. However, only S3 gave a significantly higher accuracy at the end of the session (\textbf{Fig. 2b}, \textbf{Tab S1}).  In a supplementary analysis,  we showed that this effect was not visible in the absence of feedback (\textbf{Fig. S2}).
In terms of sensitivity (correct/total MI trials), the scores were in general very high ($>83\%$) and no significant effects were reported across blocks  or strategies.

\subsection*{Motor-related spatiotemporal brain dynamics}

To understand how the brain responded to the different timing strategies, we focused on the last control block of the experiment corresponding to the best achieved accuracy in average.
First, we computed the autoregressive-based power spectrum of the EEG signals corresponding to the MI and rest trials in the four characteristic frequency bands \textit{theta} ($4-7 Hz$), \textit{alpha} ($8-12 Hz$), \textit{beta} ($13-25 Hz$) and \textit{gamma} ($26-35 Hz$)\cite{marcuse_2_2016}. 
Then, we considered the trial-averaged power spectra so to have a more robust estimation of the MI and rest condition for each subject and strategy (\textbf{Methods}).

At the group-level, all strategies exhibited a significant \textit{beta} power decrease in the MI condition as compared to the rest one, while no differences were found in the other bands (\textbf{Fig. 3a}).
In terms of spatial distribution,  all strategies involved the sensorimotor area of the brain,  while S2 further exhibited a wider extension notably including frontal premotor EEG channels in both hemispheres. 
However,  only the stronger and more focused activity of S3 translated into a significant source \textit{beta} power decrement localized in the sensorimotor area contralateral to the imagined hand grasping (\textbf{Fig. 3b}).

A more detailed time-frequency analysis of the EEG signals in the best activated channel (C1) revealed that S3 also elicited an increasingly stronger motor-related \textit{beta}-activity as compared to strategies 1 and 2 (\textbf{Fig. 4}, \textbf{Methods}). 
%
This findings indicated that performing MI during the reaching phase generates stronger brain activity responses both in space and time.  Although strategy 2 elicited higher attentional levels as measured by the pupil diameter derivative data (\textbf{Fig. S2}), that was not sufficient to generate better motor-related activation and performance.

\subsection*{Brain network changes during motor imagery}

To better understand the brain organizational properties in the different timing strategies, we performed a functional connectivity network analysis of the recorded EEG signals.
To this end, we computed the Welch-based spectral coherence in the same frequency bands considered for the power spectra (\textbf{Methods}).
The resulting brain networks consisted of nodes (the EEG channels) and weighted links measuring the amount of signal synchronization between channels.

At the group-level,  S3 elicited a higher number of motor-related functional interactions as compared to the other timing strategies. 
While this tendency was reported in every frequency band, a stronger effect was observed for the \textit{beta} band (\textbf{Fig. S3}).
To quantify how these motor-related functional connections were spatially distributed and whether they concentrated in specific brain regions, we then computed the so-called node strength which measured the total connection intensity for each node (\textbf{Methods}).

Here, only the \textit{beta}-node strength in S3 showed significant increments in the MI condition as compared to the rest one, while no differences were found in the other bands or strategies.
Notably,  the most significant EEG channels were all located over the sensorimotor area contralateral to the imagined movement and exhibited a preferential information integration with EEG channels covering prefrontal and frontal brain areas in the contralateral hemisphere and, to a less extent,  in the ipsilateral one (\textbf{Fig. 5}).

\section*{Discussion}

\subsection*{Mental imagery and intentional binding}

Mental imagery is a crucial element of noninvasive BCIs since it represents the way the users communicate their intention to the computer. However, mental imagery is a complex motor/cognitive task and it might be difficult to generate detectable brain signal features.
Maintaining a high level of engagement during the task and feeling in control of an external device significantly boosts the MI quality and elicit strong motor-related patterns. To this end, immersive setups including hybrid BCI solutions and realistic robotc effectors have effectively demonstrated their superiority compared to more traditional display-based systems. 

In such immersive setups, the role of the MI timing becomes crucial given the number of interacting components.
Notably, the period between the user's intent and the subsequent action significantly affects the sense of agency \cite{haering_was_2015}. According to intentional binding theories, when the action immediately follows the intention there is a reduced time perception by the user who feels more in control \cite{moore_intentional_2012}. Thus, in a reach-and-grasp task, deciding the moment when MI has to be performed might have important consequences on the final performance. Despite the fundamental importance of this question, the MI timing effects have been poorly investigated so far.

Here, we showed a significant difference in BCI learning depending on the timing strategy. Performing MI after the reaching phase of the robotic arm produced significantly a higher classification accuracy compared to executing the MI before or during the reaching phase. 
This result is actually inline with intentional binding phenomenon as by design the delay between the intention and the grasping action was 5, 1, and 0 s for the before, during and after strategy (\textbf{Fig. S1a}) .
These findings are particularly relevant as in most BCI protocols the MI command is inserted before the reaching phase and  provide a novel grounded criterion to boost sense of agency and therefore optimize the design of noninvasive BCIs.

\subsection*{Motor-related brain activity and embodiment}

Characterizing the spatial and temporal features of the brain signals allows to better understand the type of neural mechanisms subserving real and imagined movements.
Different types of motor-related features distinctively occur before (pre-motor potential BP), during (e.g. event-related desynchronization ERD) and also after the movement onset (e.g., event-related synchronization ERS) \cite{pfurtscheller_functional_2001}.
These temporary signal changes code the different phases of deliberate movements, both real or imagined, and typically emerge from the sensorimotor cortex within the \textit{alpha} and \textit{beta} frequency bands.
ERDs are the most used features in BCIs since they take place during the mental imagery task and can be used to directly control external applications, or be targeted by neurofeedback training with broad benefit for brain diseases but also sport activities \cite{wolpaw_brain-computer_2002}.

In highly immersive setups,  a natural question is how multimodality and the presence of external interacting devices affects the ongoing mental imagery process and more in general the motor-related brain response.
Recent results showed that including robotic effectors reinforces the sense of embodiment and leads to better  performance\cite{vavoulis_review_2023,xu_robotic_2015}. Notably, the passive observation of robot movement would be sufficient to generate ERD/ERS responses \cite{lana_erders_2013,venot_towards_2021}.
Here, we investigated how the MI motor-related brain activity were influenced by the active control of a robotic arm operating reach-and-grasp goal-oriented task.
We found that the delay between the MI control task and the robot grasping is crucial to generate spatially-focused and temporally-increasing motor-related brain responses.

The best result was obtained when the MI is performed after the robot reaching phase, immediately preceding the target grasping.  A plausible explanation would consist in the presence of a priming effect that the robot has on the ability of the brain to generate stronger task-related activity.
Indeed, recent studies demonstrated that performing MI just after motor execution and observation can have beneficial effects on the brain-related response \cite{sun_short-term_2022,daeglau_investigating_2020}.
Altogether,  these results indicated that the presence of a robot is not sufficient \textit{per-se} to create empowering embodiment effects on MI-related brain activity.
Brain responses can be too weak if MI is performed too much in advance with respect to the robot movement, or too variable if MI occurs while the robot is moving, which could induce spurious visual-related attentional processes \cite{stelmach_attentional_1994,thompson_attention_2012}. 

\subsection*{Brain connectivity networks and BCIs}

Understanding how brain areas interact beyond their mere activation is becoming more and more essential in both basic and clinical neuroscience \cite{pessoa_understanding_2014}.
Brain connectivity together with recent development in network science have allowed researchers to gain insights into the organizational and reorganizational processes of brain functioning across species, scales and conditions \cite{leeuwis_functional_2021}.
Determining which are the most connected areas, how efficiently they exchange information, or what is their tendency to form highly connected clusters, are just few example witnessing the power of these methods to quantify biologically-plausible properties such as hub formation, segregation and integration of information, and modular specialization \cite{milano_challenges_2022,koutrouli_guide_2020}.

The adoption of network-based approaches in BCI science is relatively recent, mostly due to the historical focus put by the machine learning community on the improvement of classification algorithms rather than on the identification of new brain controlling mechanisms \cite{lotte_review_2018}.
Recent works showed that sensorimotor regions can modulate their connectivity in BCI-related tasks, such as MI and resting states, and allow enriching the feature space so as to improve overall performance. 
Besides performance, the study of BCI-related brain networks can therefore provide precious information on how complex cognitive tasks integrate information from distributed regions, as in BCI learning, and inform the design of more reliable adaptive prototypes
\cite{cattai_phaseamplitude_2021, corsi_functional_2020}.

Here, we showed how the hybrid BCI control of a robotic arm in a reach-and-grasp task modulated brain connectivity networks. 
Performing MI after the reaching phase, immediately preceding the robot grasping, elicited a more connected network compared to before and during the reaching. Specifically, the contralateral sensorimotor area increased its connectivity with frontal and prefrontal areas in the same hemsiphere. These regions support intentional binding due to their involvement in higher-order cognitive functions and decision-making (pre-frontal), as well as in action prediction, temporal integration, and causal inference (frontal) \cite{cavazzana_sense_2016}.  This increased integration of information might explain the ability of sensorimotor areas to encode not only action and sensing, but also subjective agency signals, as recently demonstrated with intracortical recordings \cite{serino_sense_2022}.

\subsection*{Conclusion}

The future of noninvasive BCIs depends on their ability to become reliable and effective.
By improving the user's sense of agency and embodiment, immersive multimodal prototypes can generate more accurate brain responses and enhance performance.
To this end, we showed that optimal setups should be designed in order to maximize the user's intentional binding by placing the motor-imagery command close to the related action.

We hope that our work will stimulate further studies on the neural mechanisms underlying hybrid-BCIs beyond the effects on performance.  The related brain dynamics will be crucial to build adaptive,  interpretable and personalized BCI prototypes.
Although hybrid BCIs-brain relationships have not received much attention, we suggest that they will likely have important theoretical and practical consequences on the future generation of BCIs.


{\label{Discu_conclusion}}
\section*{Methods}
\subsection*{Experimental design}
Fifteen healthy BCI-naive subjects (8 females), aged $25 ± 1.5$ years, all right-handed, provided informed consent and participated voluntarily in the BRACCIO protocol. The protocol was approved by Inria's national ethical committee as part of the BCIPRO protocol ( authorization number 2021-35 - ref SICOERLE n°179). 
Experiments took place in the extremely controlled environment of the EEG/MEG center within the neuroimaging core facility of the Paris Brain Institute.
The experimental setup featured a robotic arm facing subjects at the opposite side of an augmented table used to display stimuli cues, visual neurofeedback, and supporting two target objects (two cans). Subjects used their gaze to select the target to reach and modulated their brain activity to determine and trigger the type of action (grasping or not) operated by a robotic arm (\textbf{Fig. 1a}, \textbf{File S1}).

The grasping action was realized by performing a sustained right hand motor imagery grasping (MI) of the right hand. The no grasping action was realized by remaining at rest.
The type of action, which was randomized within each sessions, was cued by a visual stimulus on the augmented table as a halo surrounding the cans (red for MI, blue for resting state). 
In the grasping action, the robot reaches the target,  grasps it, lifts it, puts it back into the original position and returns to its starting position.
In the no-grasping action, the robot reaches the object, remains steady and returns to its starting point.
All participants were enrolled in three consecutive experimental sessions over three weeks, each session dedicated to testing a distinct timing strategy, i.e., performing the mental task (MI or rest) before, during or after the robot arm reaching phase. The strategy order was randomized across subjects.

\subsection*{Session blocks}

 Each session was composed of four separate parts for a total duration of 3 hours in average (\textbf{Fig S6}). 
 %
First, subjects were instructed to get familiar and practice with the type of movement to be imagined in the subsequent blocks. This part consisted of 1 run comprising 10 right hand grasping trials and 10 rest trials, each lasting 11 seconds (3 seconds cue, 4 seconds task, 3 seconds end). In this block, the robot arm was not used.
In the next calibration block, the robotic hand action depended on the given cue (MI/grasping or rest/no-grasping) and was not related to the user's brain activity (feedback off). This block consisted of 3 runs of 10 MI and 10 rest trials, lasting 21.5 seconds each (6.5 seconds cue, 4 seconds task, 11 seconds end).

In the first control block (control 1), the feedback is introduced based on a LDA training of the EEG controlling features from the calibration block. The robotic hand action depended on the user's brain activity (feedback on). This block includes 3 runs of 10 MI and 10 rest trials, lasting 21.5 seconds each (6.5 seconds cue, 4 seconds task, 11 seconds end).
In the second and last control block (control 2), feedback is eventually adapted based on a new LDA training  of the EEG controlling features from control 1. The robotic hand action kept depending on the user's brain activity (feedback on). This block includes 2 runs of 10 MI and 10 rest trials, lasting 21.5 seconds each (6.5 seconds cue, 4 seconds task, 11 seconds end).

\subsection*{Hardware and software}
The robotic device was a Pollen Robotic (7 DoF) right arm mounted in front of the subject, with a claw gripper. The positions of the cans were predetermined in the robot space. Inverse Kinematics with minimum jerk was used to find optimal joints configuration for each can\cite{mick_reachy_2019}. The robot went to its target based on the choice issued for the eye tracker.

Eye-gaze was recorded and used as a command with a Tobii Pro Glasses 3 set to 50 Hz. Only the x-axis direction was used to choose between right and left. Negative values were the indication to seize the left can and positive values were the indication to seize the right can.
The augmented table consisted of a display monitor underneath a plexiglas screen to support cans. Neuro feedback and stimuli appeared beneath the cans on the black screen. The robot could reach elements of the table.

EEG signals were acquired using a 64 electrode BrainAmp system within a magnetically shielded room, with TP9 and TP10 as Reference and Ground respectively. The sampling rate was set to 500 Hz. Impedance level was set to 15 $k\Omega$ with a tolerance of 10 $k\Omega$ with ActiCap control software. The EEG acquisition was realized using the OpenViBe 3.3.0 software with BrainAmp drivers. 
Gaze flow was acquired continuously and sent to the robot server that also received decisions coming from the BCI system. The server made the synchronization between the different modalities by averaging on a time window the gaze activity (corresponding to the apparition of the crosses to select the can) and propagated the BCI order at the right moment to the robot gripper.

%

\subsection*{Online signal processing}

The eye-gaze was used to communicate to the robotic arm the target it had to reach. 
To do so, we infer the 2d gaze location from pupil positions with regards to the infrared cameras. We establish for each eye a corresponding vector pointing towards a direction, the crossing of the two eye vectors determines the estimated gaze posititon.

The EEG online data processing was performed with OpenViBE 3.3.0 \cite{renard_openvibe_2010}. The feature selection was done using HappyFeat \cite{desbois_happyfeat_2023} based on its visalization and optimization tools. 
To avoid possible transitory effects, we did not considered the first second after the stimulus cue and only kept the remaining 3 s of each trial.  In a preliminary analysis, we actually confirmed higher classification accuracies after the first second (\textbf{Fig S4}).
As a first preprocessing, common average reference (CAR) was applied to the EEG signals. The spectral analysis was carried out using the autoregressive (AR) Burg method \cite{krusienski_evaluation_2006}. The parameters chosen are the 19 for the model order, 250 ms for the windowing and 0.161 ms for the overlap, and a 1 Hz frequency bin resolution in the 4-35 Hz range, parameters were the same for all subjects.
The visual feedback was given to the subject every 250 ms by changing the halo radius proportionally to the power spectrum of the best EEG controlling feature. To allow smooth changes, we performed a logarithmic iteration consisting in buffering the power spectrum and computing its logarithm to damp its variation as in \cite{ramoser_eeg-based_1997} . 

The type of action for the robotic hand was instead calculated at the end of the trial (3 s). To this end, the average of all the accumulated power spectra were sent for a 2-class LDA classification which determined the type of action, i.e., grasping/no-grasping.

\subsection*{Offline data analysis} 

The motivation score was inferred from online questionnaires that subjects completed prior to the experimentation. Subjects answered 48 questions consisting in choosing whether to complete or not a task for a certain reward. This allowed to measure the ratio between reward and effort. Then, using a logistic regression between the actual values and those in a previously-obtained database , we computed a normalized ratio informing on the individual relative amount of motivation \cite{pessiglione_why_2018}.

Given the exceptional quality of the recorded data, no artefact rejection procedure was applied to the EEG signals for the offline analysis.
For the power spectra estimation, we used the same parameters as in the online analysis, but we then averaged all the segments and trials so to have a characteristic MI and rest power spectrum for each subject and strategy.
Similarly, characteristic EEG time-frequency maps were obtained by computing the AR-based power spectra with the following parameters: temporal resolution of 0.011s and frequency resolution of 0.1 Hz, the Burg parameters were set to 250 ms for time window with 100 ms of overlap and a filter order set to 19 following \cite{bufalari_autoregressive_2006}.

The computation of the source space was done in two steps. In a first step, we computed a forward model from the sensor space to a template brain based on a combination of 40 healthy MRI scans from freesurfer software with 20484 vertices. In a second step, we project the sensor activity onto the source space. The regularisation parameter $\lambda^2$ is set to $1/3$ for source estimation
following\cite{vallarino_tuning_2023}
using the weighted minimum norm method\cite{grech_review_2008}. We estimated the power spectrum in the source space using multitaper method with an overlap of 0.5s corresponding to the degree of overlap between the tapered time segments with Hann window. 

Functional connectivity was computed from the segmented EEG signals by means of the Welch-based spectral coherence \cite{cattai_phaseamplitude_2021}. The main parameters were: window length 500 ms, overlap 250 ms and frequency resolution 1 Hz with a Hanning window.
The resulting networks consisted of connectivity matrices where the $i,j$ entry contained the amount of signal synchronization between the EEG channel $i$ and $j$ at a given frequency bin. 
From these networks, we further computed the node strength as the sum of all the connectivity matrices' rows \cite{gonzalez-astudillo_network-based_2021}. Node strengths allow to have quantitative measure of the amount of total connectivity of each EEG channel.
Both connectivity matrices and node strengths, were finally averaged across segments and trials so to have characteristic values for each subject, task (MI, rest) and strategy.

To investigate the group-level statistical difference between MI and rest tasks, we used a paired cluster-based permutation t-test applied to the motor-related score given by the relative difference between tasks

\begin{equation}
\frac{MI-rest}{rest}=\frac{MI}{rest}-1
\end{equation} 

This normalizing transformation was applied to every considered metric, i.e. power, time-frequency power, functional connectivity and node strength. The maximal statistical threshold level was set to $\alpha = 0.05$.

\begin{small}
\bibliography{JournalPaperThesis.bib}
\end{small}
\clearpage
\section*{Acknowledgements}

We would like to thank Will Hopper and Jean Daunizeau for having provided the code to compute the motivation score and for insightful discussion. 
FDVF acknowledges support from the European Research Council (ERC), Grant Agreement No. 864729 and from the Agence Innovation Defense DGA.
All data needed to evaluate the conclusions in the paper are present in the paper and/or the Supplementary Materials.
The data (power spectra, time-frequency, functional connectivity, pupil diameter variations, motivation scores) can be provided by Fabrizio, De Vico Fallani pending scientific review and a completed material transfer agreement. Requests for these data should be submitted to fabrizio-de-vico-fallani@inria.fr
The authors declare that they have no competing interests.

\clearpage
\begin{figure}
    \centering
    \includegraphics[width=\textwidth]{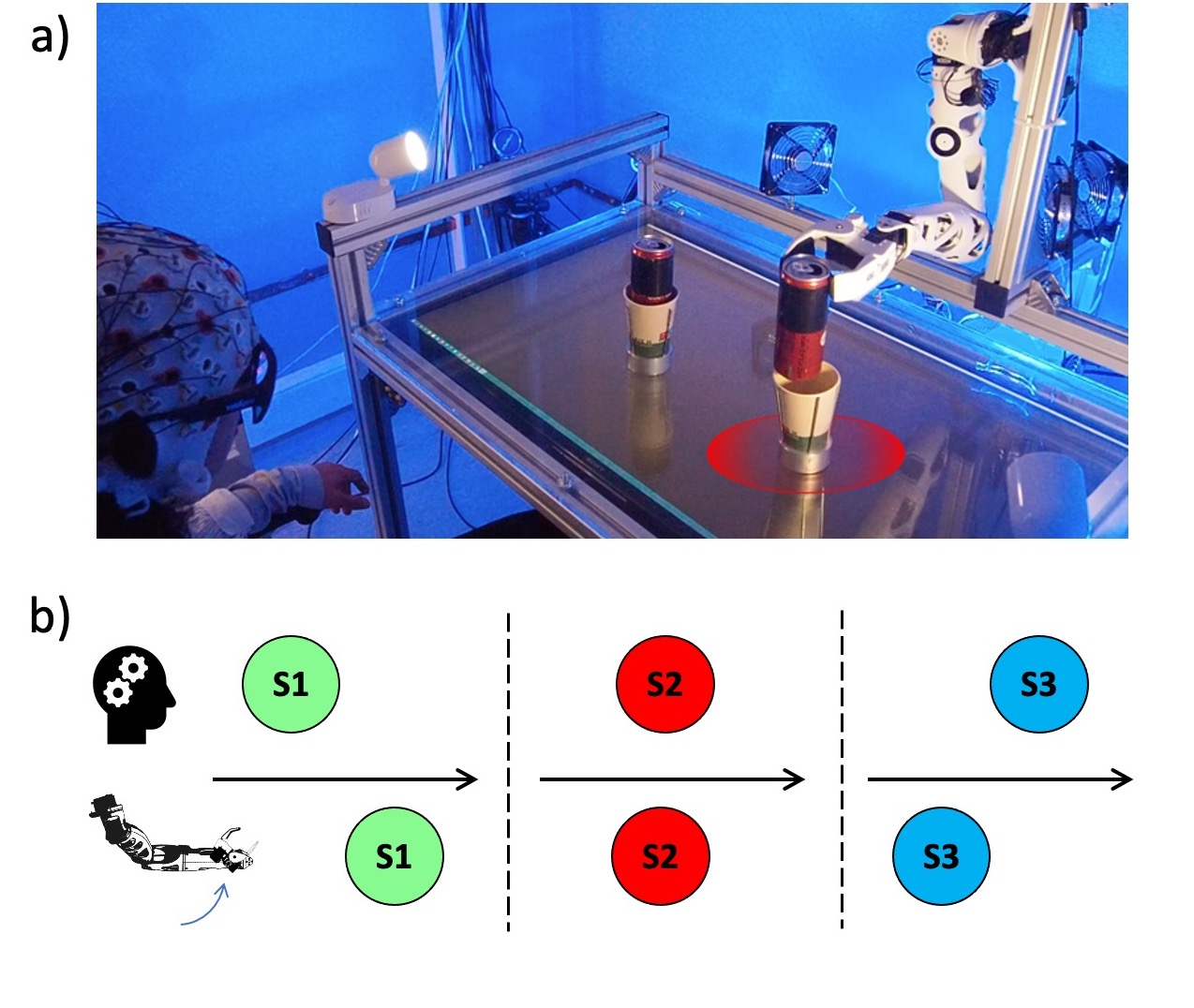}
    \caption{\textbf{Structure of the experimental setup}.
\\
\textbf{a)} Hybrid-BCI system composed of a 64 EEG BrainAmp device, the Tobii Pro Glasses 3 Eyetracker, the augmented table with the display monitor underneath the glass and a Pollen Robotics's Reachy arm facing the subject. The user selects the target via the gaze position. The position is sent to the robot which reaches the target.  Once the robot reaches the target, its grasping depends on the user’s motor imagery (MI) activity command. 
\\
\textbf{b) }The three different timings of control investigated in the protocol. Strategy 1 (in green) - performing MI before the robot reaches the target can. Strategy 2 (in red) - performing MI while the robot reaches the target, strategy 3 (in blue) - performing MI after the robot has reached the target.  Among all strategies, the latter is the one temporally closest to the subsequent grasping robot action (Fig. S1a).
}
    \label{fig:1}
\end{figure}

\clearpage

\begin{figure}
    \centering
    \includegraphics[width=\textwidth,trim={1cm 0 1cm 0}]{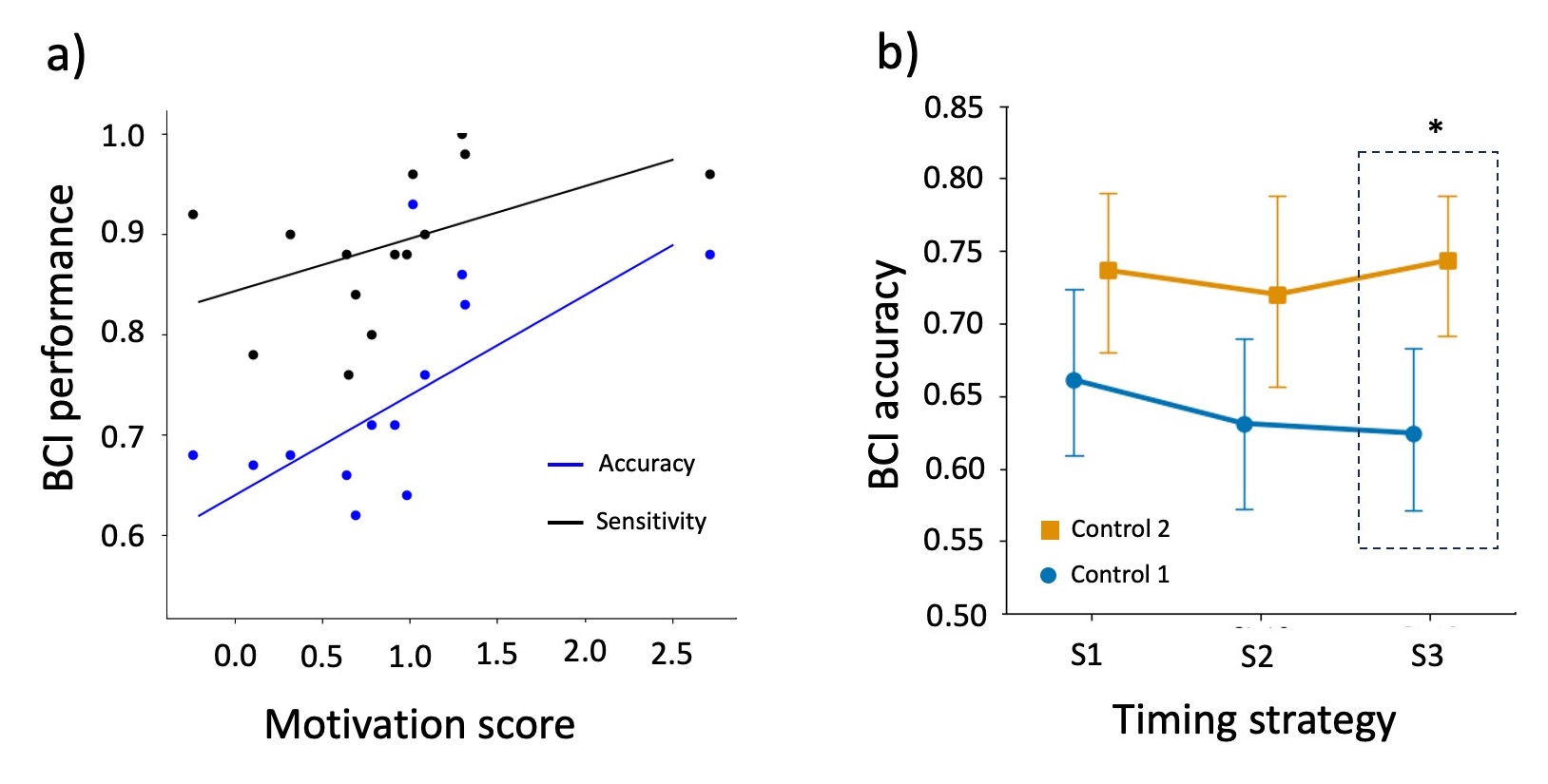}
    \caption{\textbf{Participants’ motivation and BCI performance}. 
    \\
\textbf{a)} Correlation between the highest performance of subjects across strategies in terms of accuracy and sensitivity as a function of their motivation as measured by the Reward/Effort test \cite{pessiglione_why_2018}.  Spearman correlation: Sensitivity in black $R=0.595$, $p_{val}=0.024$. Accuracy in blue $R=0.683$,$p_{val}=0.007$. 
\\
\textbf{b)} Group-averaged accuracy in the two control blocks for each timing strategy . 2-way ANOVA reveals a significant learning effect (Block F=14.861523
 $p<0.0002$), while there are no significant effects for the timing strategy. A post-hoc analysis with Bonferroni correction indicates that only strategy 3 leads to a significant increase of classification accuracy at the end of the session ($F=4.9837$, $p_{corrected} = 0.003$). Control block 1 in blue, control block 2 in orange. Asterisk denotes statistically significant difference with $\alpha=0.05$ threshold.
}
    \label{fig:2}
\end{figure}

\clearpage

\begin{figure}
    \centering
    \includegraphics[width=\textwidth,trim={1cm 0 1cm 0}]{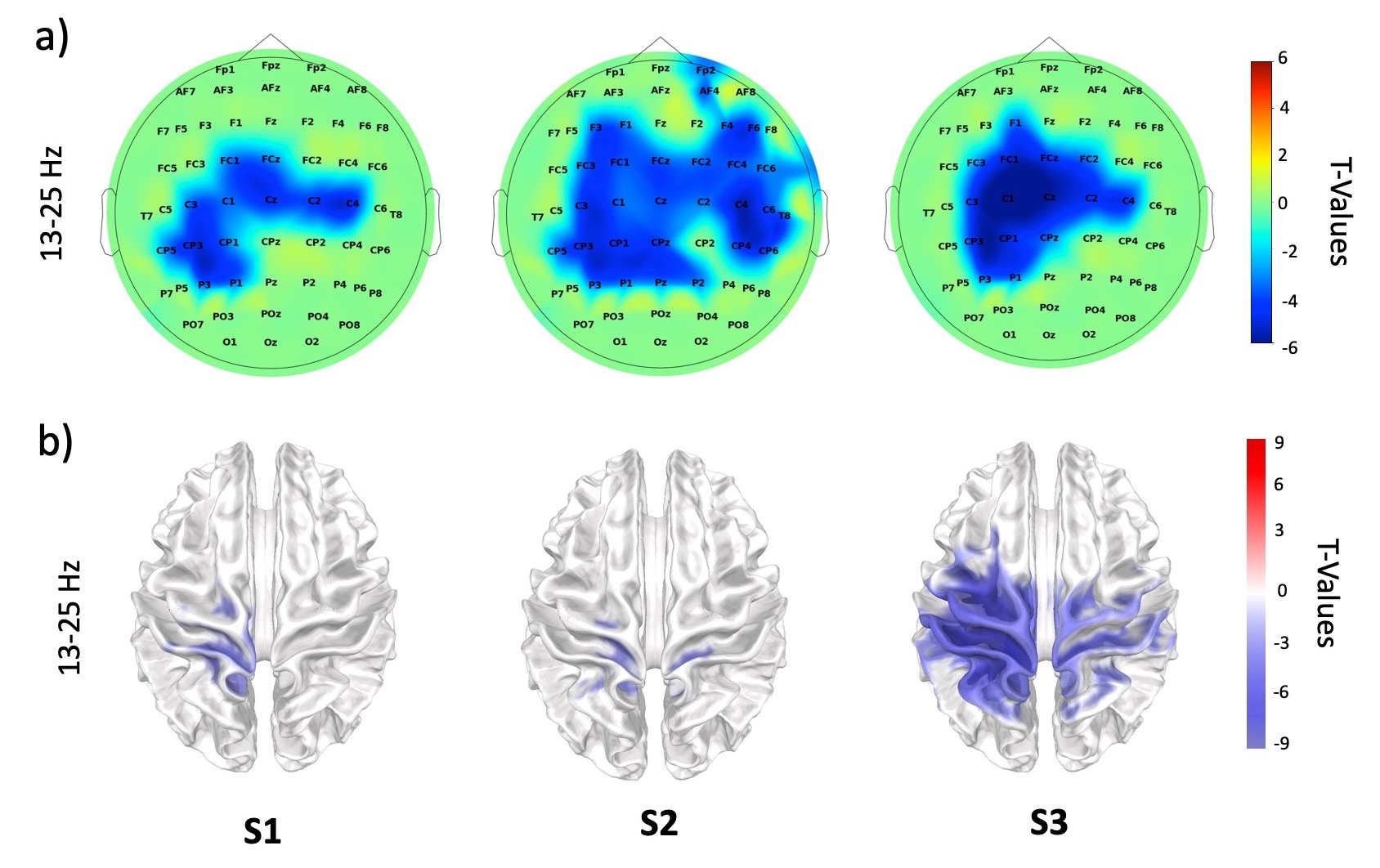}
    \caption{\textbf{Maps of motor-related brain activity for different MI timing strategies}. 
    \\
\textbf{a)} Sensor space activity. Cluster-based permutation T-test on the relative difference between the average power spectra of MI and Rest trials across subjects (Methods).  Data refer to the Control block 2 and \textit{beta} frequency band (13-25Hz).  Only T-values corresponding to $p<0.05$ are shown. The distance matrix for the cluster computation is set considering a threshold of 40 mm between the EEG channels (based on the average distance between electrodes in a 10-10 international system \cite{seeck_standardized_2017}). The cluster forming threshold is computed using percent point function at $\alpha = 0.05$ for $n=15$. 
\\
\textbf{b)} Source space activity. Cluster-based permutation T-test on the relative difference between the average power spectra of MI and Rest trials across subjects (Methods).  Data refer to the Control block 2 and \textit{beta} frequency band (13-25Hz). Source space dipoles’ signals are estimated using a weight minimum norm estimation (wMNE) and a average head model (Methods). The distance matrix for the cluster computation is set considering a threshold of 20 mm between the source dipoles (based on the average distance between dipoles in a forward model\cite{song_eeg_2015}). The cluster forming threshold is computed using percent point function at $\alpha = 0.05$ for $n=15$. Only strategy 3 present significant T-values ($p<0.05$), while the weaker activations in the other two other strategies are shown only for illustrative purposes.
}
    \label{fig:3}
\end{figure}
\clearpage
\begin{figure}
    \centering
    \includegraphics[width=\textwidth,trim={1cm 0 1cm 0}]{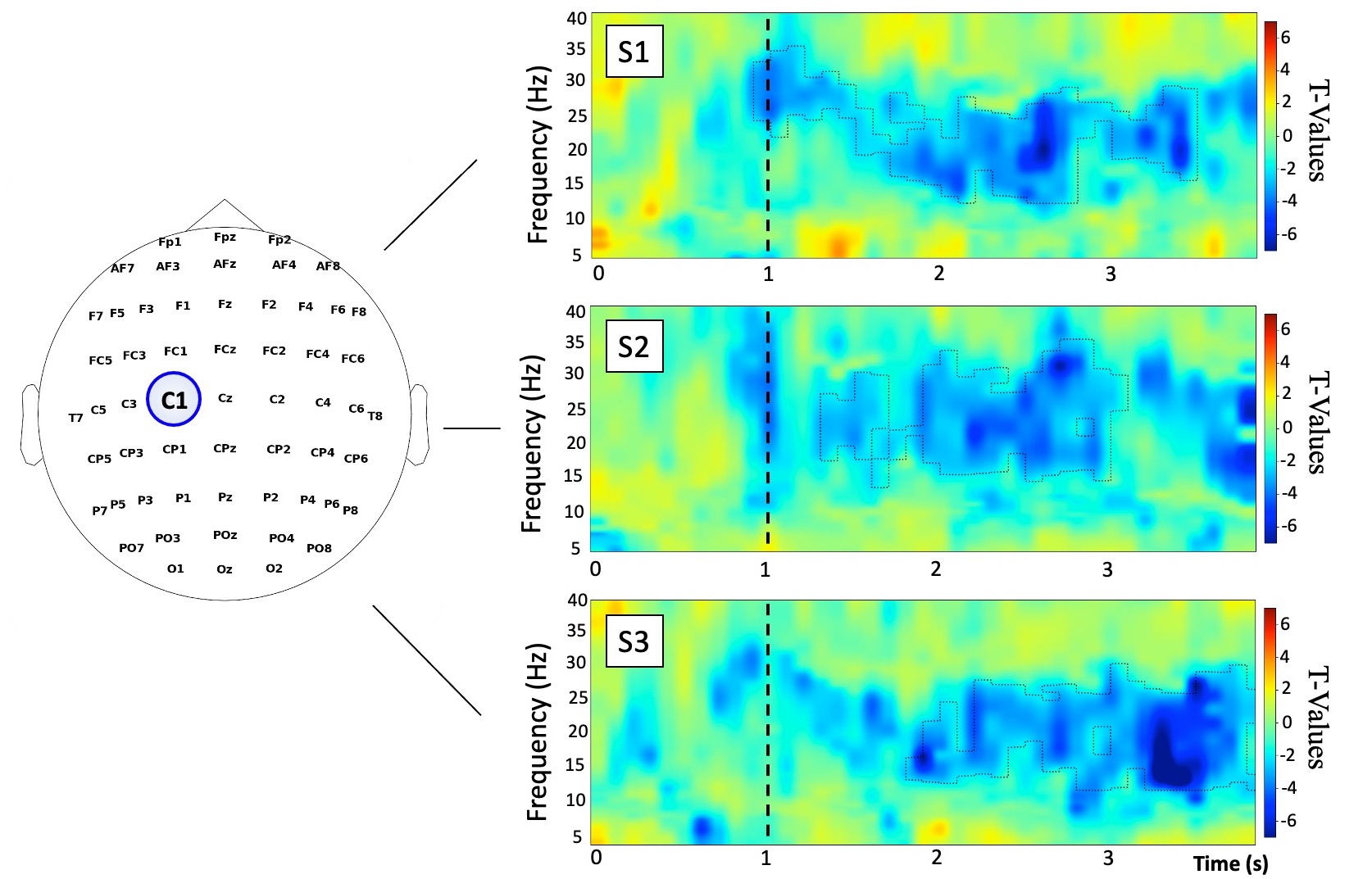}
    \caption{\textbf{Time-frequency maps of motor-related activity for different MI timing strategies}
Cluster-based permutation T-test on the relative difference between the time-varying average power spectra of MI and Rest trials across subjects (Methods).  Data refer to the Control block 2 and are computed with an autoregressive model (Burg) on a sliding window of 0.011 s and frequency resolution 0.1 Hz (Methods). T-values corresponding to $p<0.05$ are highlighted by dashed contours. The distance matrix for the cluster computation at each point is given by closest pixels  \cite{brinkman_distinct_2014}. The cluster forming threshold is computed using percent point function at $\alpha = 0.05$ for $n=15$. Vertical dashed lines indicate that significant motor-related activity starts to appear after 1 s from the cue.
}
    \label{fig:4}
\end{figure}
\clearpage
\begin{figure}
    \centering
    \includegraphics[width=\textwidth,trim={1cm 0 1cm 0}]{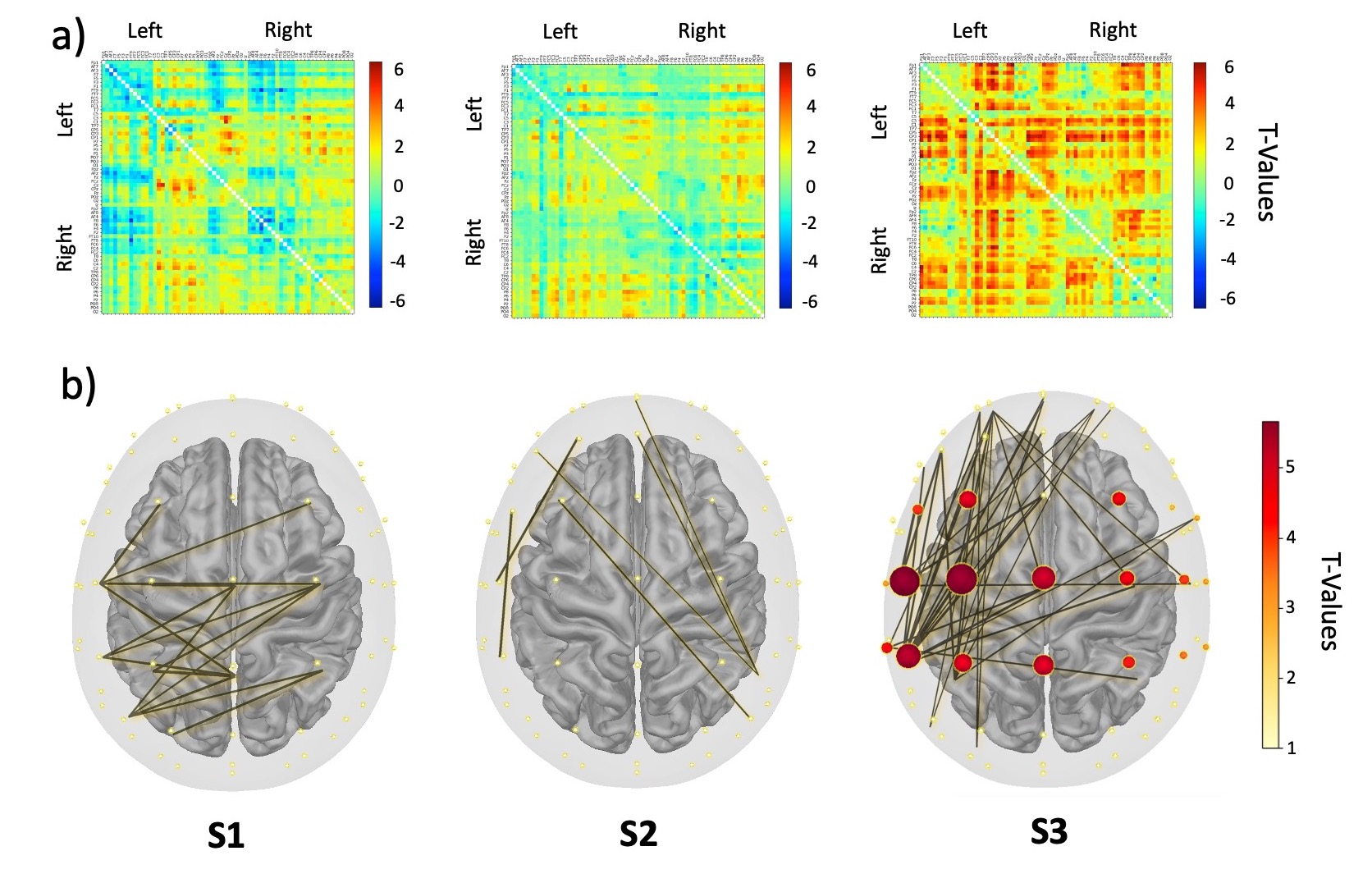}
    \caption{\textbf{Motor-related brain connectivity networks for different MI timing strategies}. 
    \\
Statistical connectivity matrices. Each entry corresponds to an EEG channel pair. Left=left hemisphere. Right=right hemisphere. The color codes for the permutation T-test value on the relative difference between the average spectral beta (13-25 Hz) coherence in the MI and Rest trials across subjects (Methods).
Cluster based permutation T-test values on the relative difference of the node strength between average MI and Rest trials across subjects (Methods). Results refer to the beta band (13-25Hz) in last session block (Control 2). Nodes correspond to EEG channels. The larger and darker is the node the higher is the associated T-value. Only T-values corresponding to $p<0.05$ are shown. The most significant nodes in the sensorimotor related zone are CP3,C1, and C3. The distance matrix for the cluster computation is set considering a threshold of 40 mm between the EEG nodes \cite{seeck_standardized_2017}. The cluster forming threshold is computed using percent point function at $\alpha = 0.05$ for $n=15$. Links are shown for illustrative purposes and correspond to the strongest T-values from the related connectivity matrices in a), ie $p<0.001$ for strategy 1 and 2, $p<10e-06$ for strategy 3.}
    \label{fig:5}
\end{figure}

\clearpage
\end{document}